\documentclass[journal]{IEEEtran}

\usepackage[utf8]{inputenc}
\usepackage{graphicx,color,epsf}
\usepackage{amsmath,amsfonts,amssymb,amscd,bm}
\usepackage{dsfont}
\usepackage{flushend}                                       
\usepackage{struktex}

\usepackage{subcaption}
\usepackage[font=footnotesize]{caption}
\usepackage{mathtools}
\usepackage{siunitx}

\usepackage{bbm}
\usepackage{xcolor}
\usepackage{algorithm}
\usepackage[acronym]{glossaries}
\usepackage{xspace}
\usepackage{algpseudocode}
\usepackage{upgreek}
\usepackage{hyperref}
\usepackage[capitalise]{cleveref}                           
\usepackage{eso-pic}

\AddToShipoutPicture*{\footnotesize\sffamily\raisebox{0.8cm}{\hspace{1.65cm}\fbox{\parbox{\textwidth}{\copyright~2019 IEEE. Personal use of this material is permitted. Permission from IEEE must be obtained for all other uses, in any current or future media, including reprinting/republishing this material for advertising or promotional purposes, creating new collective works, for resale or redistribution to servers or lists, or reuse of any copyrighted component of this work in other works.}}}}

\DeclareMathAlphabet{\pazocal}{OMS}{zplm}{m}{n}

\crefname{equation}{}{}                                     
\Crefname{equation}{Equation}{Equations}                    

\newacronym{FIT}{\protect FIT}{finite integration technique}
\newacronym{PGD}{\protect PGD}{proper generalized decomposition}

\catcode`@=11
\newcommand{\frownfill}{\ensuremath{\scriptscriptstyle\m@th\mathord\frown}}
\newcommand{\bow}[1]{\ensuremath{\vbox{\m@th\ialign{##\crcr
      \hfil\frownfill\hfil\crcr\noalign{\kern-0.2\p@\nointerlineskip}
      $\hfil\displaystyle{#1}\hfil$\crcr}}}}
\newcommand{\bbow}[1]{\ensuremath{\vbox{\m@th\ialign{##\crcr
     \hfil\frownfill\hfil\crcr\noalign{\kern-0.7\p@\nointerlineskip}
     \hfil\frownfill\hfil\crcr\noalign{\kern-0.3\p@\nointerlineskip}
      $\hfil\displaystyle{#1}\hfil$\crcr}}}}
\newcommand{\bbbow}[1]{\ensuremath{\vbox{\m@th\ialign{##\crcr
     \hfil\frownfill\hfil\crcr\noalign{\kern-0.7\p@\nointerlineskip}
     \hfil\frownfill\hfil\crcr\noalign{\kern-0.7\p@\nointerlineskip}
     \hfil\frownfill\hfil\crcr\noalign{\kern-0.3\p@\nointerlineskip}
      $\hfil\displaystyle{#1}\hfil$\crcr}}}}
\newcommand{\widefrownfill}{\ensuremath{\m@th\mathord\frown}}
\newcommand{\widebow}[1]{\ensuremath{\vbox{\m@th\ialign{##\crcr
      \hfil\widefrownfill\hfil\crcr\noalign{\kern-0.9\p@\nointerlineskip}
      $\hfil\displaystyle{#1}\hfil$\crcr}}}}
\newcommand{\widebbow}[1]{\ensuremath{\vbox{\m@th\ialign{##\crcr
     \hfil\widefrownfill\hfil\crcr\noalign{\kern-1.8\p@\nointerlineskip}
     \hfil\widefrownfill\hfil\crcr\noalign{\kern-0.9\p@\nointerlineskip}
      $\hfil\displaystyle{#1}\hfil$\crcr}}}}
\newcommand{\widebbbow}[1]{\ensuremath{\vbox{\m@th\ialign{##\crcr
     \hfil\widefrownfill\hfil\crcr\noalign{\kern-1.8\p@\nointerlineskip}
     \hfil\widefrownfill\hfil\crcr\noalign{\kern-1.8\p@\nointerlineskip}
     \hfil\widefrownfill\hfil\crcr\noalign{\kern-0.9\p@\nointerlineskip}
      $\hfil\displaystyle{#1}\hfil$\crcr}}}}

\begin{document}
\title{\fontsize{17}{20}\bf
Proper Generalized Decomposition of Parameterized Electrothermal Problems Discretized by the Finite Integration Technique
}
\author{
\IEEEauthorblockN{Alexander Krimm\textsuperscript{1}, Thorben Casper\textsuperscript{1,2}, Sebastian Schöps\textsuperscript{1,2}, Herbert De Gersem\textsuperscript{1,2}, Ludovic Chamoin\textsuperscript{3}}
  \\
  {\textsuperscript{1}\small Institut für Theorie Elektromagnetischer Felder, Technische Universität Darmstadt, Darmstadt, Germany}
  \\
  {\textsuperscript{2}\small Graduate School of Computational Engineering, Technische Universität Darmstadt, Darmstadt, Germany}
  \\
  {\textsuperscript{3}\small Laboratory of Mechanics and Technology, ENS Cachan, CNRS, Université Paris-Saclay, Cachan, France\vspace{-1em}}
}
\maketitle

\begin{abstract}
    The proper generalized decomposition is applied to a static electrothermal model subject to uncertainties.
    A reduced model that circumvents the curse of dimensionality is obtained.
    The quadratic electrothermal coupling term is non-standard and requires the introduction of a trilinear form.
    An existing finite integration technique based solver is used to demonstrate the opportunities and difficulties in integrating the proper generalized decomposition in existing codes.
\end{abstract}
 \section{Introduction}

A recurring problem for simulation tools in engineering is the high number of possible parameter combinations.
For optimization and uncertainty quantification, surrogate models are required.
Classical techniques construct surrogate models from empirical knowledge with a subsequent fitting to measurement data.
However, such simplified models are usually only valid in a limited parameter range.

\Gls*{PGD} allows to obtain a reduced model for a wide class of problems~\cite{Chinesta_2013ab,Ammar_2007aa}.
In contrast to other techniques, its complexity is not subject to the curse of dimensionality.
It separates the high dimensional solution into univariate functions using an iterative algorithm.
Whereas mainly used in mechanics, few works of \gls*{PGD} in computational electromagnetics have been reported on,~e.g., for extracting parasitic capacitances of interconnects in integrated circuits~\cite{Li_2017ab} and for magnetoquasistatic field-circuit couplings~\cite{Henneron_2015aa}.
Further examples include the application to skin effect problems~\cite{Pineda-Sanchez_2010aa} and the use for space time separation of magnetothermal problems~\cite{Qin_2016aa}.
However, to the best of our knowledge, the \gls*{PGD} has not been applied to an electrothermally coupled problem.
Since it is desired to combine the \gls*{PGD} technique with existing solvers and frameworks, minimally intrusive implementations are a current research topic as well~\cite{Giraldi_2015aa}.

\begin{figure}
    \centering
    \includegraphics{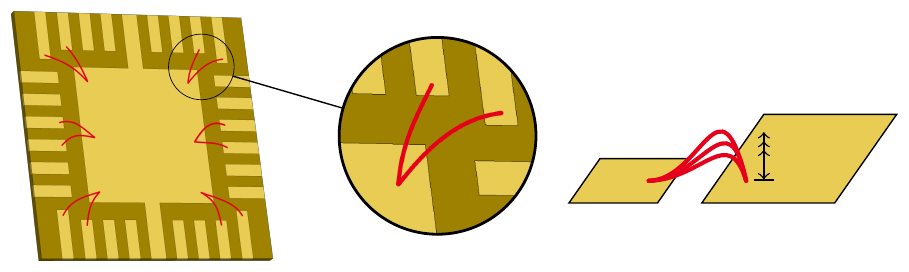}
        \caption{Microelectronic chip package with substrate (dark yellow), conducting regions (yellow) and bond wires (red) of uncertain geometry.}
    \label{fig:chip}
\end{figure}

In this paper, we exploit \gls*{PGD} for reducing the electrothermal model of a microelectronic chip package, see \cref{fig:chip}.
Such a package typically contains a high number of bond wires affected by variability in the fabrication process.
Any geometric uncertainties influence the electrical and thermal behavior of the package and may determine the lifespan of the device.
Evaluating this problem for all parameter combinations exceeds the capability of classical techniques as, e.g., Monte Carlo simulation.
The construction of a surrogate model as described in this paper can also be used for hybrid approaches as, e.g., reported on in \cite{Casper_2016ah}.

The paper is structured as follows.
In \cref{sec:problem}, we introduce the parametric electrothermal problem.
The \gls*{PGD} algorithm applied to the electrothermal coupling and its usage with the \gls*{FIT} is presented in \cref{sec:pgd}.
For illustration and validation purposes, in \cref{sec:results}, we show numerical results when solving a 1D model problem and a 3D microelectronic chip package using \gls*{PGD}.
Finally, \cref{sec:conclusion} concludes the paper.
 \section{Parameterized Electrothermal Problem}
\label{sec:problem}

Whereas the area of applications for \gls*{PGD} is vast, the focus in this paper is the static electrothermal problem.
We combine the electrokinetic problem on the spatial domain $\Omega$ of dimension $d$ obtained from \textsc{Maxwell}'s equations~\cite{Jackson_1998aa} with the static heat equation~\cite{Widder_1975aa} to obtain
\begin{subequations}
    \begin{alignat}{2}
    -\nabla \cdot\left(\sigma(\mathbf{x};\ensuremath{\boldsymbol{\upmu}}\xspace)\nabla\ensuremath{\varphi}\xspace(\mathbf{x};\ensuremath{\boldsymbol{\upmu}}\xspace)\right)&=0,
    \quad &&\mathbf{x}\in\Omega,\\
    -\nabla \cdot\left(\lambda(\mathbf{x};\ensuremath{\boldsymbol{\upmu}}\xspace)\nabla T(\mathbf{x};\ensuremath{\boldsymbol{\upmu}}\xspace)\right)&=Q(\sigma,\ensuremath{\varphi}\xspace),
    \quad&&\mathbf{x}\in\Omega,
    \end{alignat}
    \label{eq:etspaceonly}
\end{subequations}
with suitable boundary conditions.
Here, $\ensuremath{\varphi}\xspace(\mathbf{x};\ensuremath{\boldsymbol{\upmu}}\xspace)$ is the electric potential, $T(\mathbf{x};\ensuremath{\boldsymbol{\upmu}}\xspace)$ the temperature, $\sigma(\mathbf{x};\ensuremath{\boldsymbol{\upmu}}\xspace)$ is the electric conductivity, $\lambda(\mathbf{x};\ensuremath{\boldsymbol{\upmu}}\xspace)$ is the thermal conductivity and $Q(\sigma,\ensuremath{\varphi}\xspace):=\sigma\lvert\nabla\ensuremath{\varphi}\xspace\rvert^{2}$ describes the resistive losses.
The parameters $\ensuremath{\boldsymbol{\upmu}}\xspace\in\Psi:=\times_{p=1}^{n_{\mu}}I_{p}$, express variations due to the fabrication process, where $n_{\mu}$ is the number of parameters and $I_{p}\in\mathds{R}$ the domain of parameter $p$.
These variations can, e.g., arise due to an uncertain length of the bond wires in a microelectronic chip package, see \cref{fig:chip}.
Assuming temperature independent material properties, we obtain a one-directional coupling by the resistive losses $Q$.

In the following, we treat problem~\cref{eq:etspaceonly} as a high dimensional problem by introducing the space-parameter variable
\begin{align}
    \tilde{\mathbf{x}}&=\begin{pmatrix}\mathbf{x},\mu_{1},\ldots,\mu_{n_{\mu}}\end{pmatrix}^{\top}\in\Theta:=\Omega\times\Psi,
\end{align}
such that~\cref{eq:etspaceonly} simplifies to
\begin{subequations}
    \begin{alignat}{2}
        -\nabla \cdot\left(\sigma(\tilde{\mathbf{x}})\nabla_{\!\mathbf{x}}\ensuremath{\varphi}\xspace(\tilde{\mathbf{x}})\right)&=0,\quad &&\tilde{\mathbf{x}}\in\Theta,\label{eq:electrokinetic}\\
        -\nabla \cdot\left(\lambda(\tilde{\mathbf{x}})\nabla_{\!\mathbf{x}}T(\tilde{\mathbf{x}})\right)&=Q(\sigma,\ensuremath{\varphi}\xspace),\quad&&\tilde{\mathbf{x}}\in\Theta,\label{eq:staticthermal}
    \end{alignat}
    \label{eq:etprb}
\end{subequations}
where $\nabla_{\!\mathbf{x}}$ is the gradient in $\mathbf{x}$-direction. \section{Proper Generalized Decomposition}
\label{sec:pgd}

To apply the \gls*{PGD} to \cref{eq:etprb}, we recognize that the subproblems \cref{eq:electrokinetic,eq:staticthermal} exhibit an identical mathematical structure except for the different right-hand sides.
First, we focus on the left-hand side using \cref{eq:electrokinetic} as an example and secondly, we consider the electrothermal coupling term on the right-hand side of \cref{eq:staticthermal}.
Since the \gls*{PGD} formulation is typically based on the weak formulation~\cite{Chinesta_2013ab}, we introduce test functions $v$ and use integration by parts.
Whereas the formulation here is done for an arbitrary number of parameters and becomes quite technical, we show a simplified 1D example with one parameter in \cref{sec:1D}.

\subsection{Electrokinetic Problem}\label{sec:pgdelectric}
The weak form of \cref{eq:electrokinetic} reads: find $\ensuremath{\varphi}\xspace\in{}V_{\Theta}$ s.t.
\begin{align}
    \langle\sigma(\tilde{\mathbf{x}})\nabla_{\!\mathbf{x}}\ensuremath{\varphi}\xspace(\tilde{\mathbf{x}}),\nabla{}v(\tilde{\mathbf{x}})\rangle_{V_{\Theta}}&=0,\quad\forall{}v\in{}V_{\Theta},
    \label{eq:estatic_weak}
\end{align}
where $\langle u,v\rangle_{V}$ denotes the $L^{2}$ inner product of functions $u$ and $v$ in $V$ and $V_{\Theta}:=\otimes_{p=1}^{n_{p}}V_{p}$ with $V_{1}=H^{1}(\Omega)$ and $V_{p}=L^{2}(I_{p})$ for $p=2,\ldots,n_{p}$ and $n_{p}=n_{\mu}+1$, representing one spatial and $n_{\mu}$ parameter spaces.
Following the \gls*{PGD} approach~\cite{Chinesta_2013ab}, we assume that the solution can be approximated by
\begin{align}
    \ensuremath{\varphi}\xspace(\tilde{\mathbf{x}})&\approx\ensuremath{\varphi}\xspace_{\text{PGD}}^{m}(\tilde{\mathbf{x}}):=\sum_{s=1}^{m}u^{s}(\tilde{\mathbf{x}}):=\sum_{s=1}^{m}\prod_{p=1}^{n_{p}}u_{p}^{s}(\tilde{\mathbf{x}}_{p})
    \label{eq:trialfunction}
\end{align}
where $m$ is the number of \gls*{PGD} modes.
Note that we do not separate the spatial dependencies to enable the usage of existing 3D solvers.
Therefore, we have $\tilde{\mathbf{x}}_{1}=\mathbf{x}\in\mathds{R}^{3}$ and $\tilde{\mathbf{x}}_{p}\in\mathds{R}$ for $p>1$.
The idea of the \gls*{PGD} is the separation of variables such that a decomposition of all quantities in~\cref{eq:estatic_weak} is required.
The electric conductivity is decomposed as
\begin{align}
    \sigma(\tilde{\mathbf{x}})=\sum_{q=1}^{n_{\sigma}}\sigma_{q}(\tilde{\mathbf{x}})&:=\sum_{q=1}^{n_{\sigma}}\prod_{p=1}^{n_{p}}\sigma_{q,p}(\tilde{\mathbf{x}}_{p}).
\end{align}
which is trivial for, e.g., piecewise constant parameterized coefficients on $n_{\sigma}$ different subdomains of the problem.
For more complex cases, appropriate approximations may be used to reduce the required number of terms $n_{\sigma}$.
The resulting decomposition of \cref{eq:estatic_weak} reads,
\begin{align}
    \sum_{q=1}^{n_{\sigma}}\sum_{s=1}^{m}\underbrace{\langle\sigma_{q}(\tilde{\mathbf{x}})\nabla_{\!\mathbf{x}}\prod_{p=1}^{n_{p}}u_{p}^{s}(\tilde{\mathbf{x}}_{p}),\nabla_{\!\mathbf{x}}{}v(\tilde{\mathbf{x}})\rangle_{V_{\Theta}}}_{=:\mathcal{L}(q,s)}=0,
    \label{eq:weakpgdprb}
\end{align}
where the abbreviation $\mathcal{L}$ was introduced for conciseness.

To solve \cref{eq:weakpgdprb}, we assume that all modes $s<m$ are known and solve
\begin{align}
    \sum_{q=1}^{n_{\sigma}}\mathcal{L}(q,m)=-\sum_{q=1}^{n_{\sigma}}\sum_{s=1}^{m-1}\mathcal{L}(q,s)
    \label{eq:estatic_pgd}
\end{align}
for mode $m$.
Starting with $m=0$, where the right-hand side is zero, an iterative enrichment scheme is established.
Let $p^{*}\in\{1,\ldots,n_{p}\}$ be an arbitrary direction, then this nonlinear equation can be solved with a fixed point iteration solving for the $p^{*}$-th direction in each step~\cite{Ammar_2014aa}.
Accordingly, the test function for mode $m$ is separated as
\begin{align}
    v^{m}(\tilde{\mathbf{x}})&=\sum_{p^{*}=1}^{n_{p}}v_{p^{*}}(\tilde{\mathbf{x}}_{p^{*}})\prod_{\substack{p=1\\p\neq{}p^{*}}}^{n_{p}}u_{p}^{m}(\tilde{\mathbf{x}}_{p}),
    \label{eq:testfunctions}
\end{align}
where the \gls*{PGD} suggests to choose the already calculated functions $u_{p}^{m}$ as test functions for the directions $p\neq{}p^{*}$.
In the following, we omit the parametric dependencies for a compact notation.
All parameter functions $u_{p}^{m}$ with $p\ne{}p^{*}$ are considered known, giving
\begin{align}
    \mathcal{L}(q,m)=\left\langle\sigma_{q}\nabla_{\!\mathbf{x}}\left(u_{p^{*}}^{m}\prod_{\substack{p=1\\p\neq{}p^{*}}}^{n_{p}}u_{p}^{m}\right),\nabla_{\!\mathbf{x}}\left(v_{p^{*}}\prod_{\substack{p=1\\p\neq{}p^{*}}}^{n_{p}}u_{p}^{m}\right)\right\rangle_{V_{\Theta}}.
    \label{eq:LqmFixedPoint}
\end{align}
A further simplification of $\mathcal{L}$ depends on whether the current step solves for the spatial solution or for a parametric solution.
In the following, we outline the spatial case $p^{*}=1$ and leave the parametric case to the reader. 
If the current step evaluates the spatial problem, we obtain
\begin{align}
    \mathcal{L}(q,m)=\left\langle\sigma_{q,1}\nabla_{\!\mathbf{x}}u_{1}^{m},\nabla_{\!\mathbf{x}}{}v_{1}\right\rangle_{H^{1}(\Omega)}
    \prod_{p=2}^{n_{p}}\left\langle{}\sigma_{q,p}u_{p}^{m},u_{p}^{m}\right\rangle_{L^{2}(I_{p})},
\end{align}
which can also be written as
\begin{align}
    \mathcal{L}(q,m)=a_{q,1}(u^{m}_{1},v_{1})\prod_{p=2}^{n_{p}}a_{q,p}(u^{m}_{p},u_{p}^{m}),
    \label{eq:estatic_fp_step}
\end{align}
with the bilinear forms defined as
\begin{align}
    a_{q,1}(u,v)&:=\langle\sigma_{q,1}\nabla_{\!\mathbf{x}}u,\nabla_{\!\mathbf{x}}v\rangle_{H^{1}(\Omega)},\\
    a_{q,p}(u,v)&:=\langle\sigma_{q,p}u,v\rangle_{L^{2}(I_{p})}.
\end{align}

Applying a standard \textsc{Galerkin} finite element discretization to~\cref{eq:estatic_pgd} with $\mathcal{L}$ as given in~\cref{eq:estatic_fp_step}, the linear system for step $k$ of the fixed point iteration reads
\begin{align}
    \sum_{q=1}^{n_{\sigma}}\mathbf{M}_{q,1}\mathbf{u}_{1}^{m,k}\prod_{p=2}^{n_{p}}\alpha_{q,p}^{m,m,k}
    &=-\sum_{q=1}^{n_{\sigma}}\sum_{s=1}^{m-1}\mathbf{M}_{q,1}\mathbf{u}_{1}^{s,k}\prod_{p=2}^{n_{p}}\alpha_{q,p}^{m,s,k},
    \label{eq:PGDsystemDisc}
\end{align}
where we have introduced the scalar constants
\begin{align}
    \alpha_{q,p}^{i,j,k}&:=(\mathbf{u}_{p}^{i,k})^{\top}\mathbf{M}_{q,p}\mathbf{u}_{p}^{j,k},
\end{align}
with the integers $i$ and $j$, and the finite element matrices $\mathbf{M}_{q,1}$ and $\mathbf{M}_{q,p}$ stemming from the discretization of the bilinear forms $a_{q,1}(u,v)$ and $a_{q,p}(u,v)$, respectively, apart from the corresponding weighting vectors.

\begin{figure}
    \centering
    \input{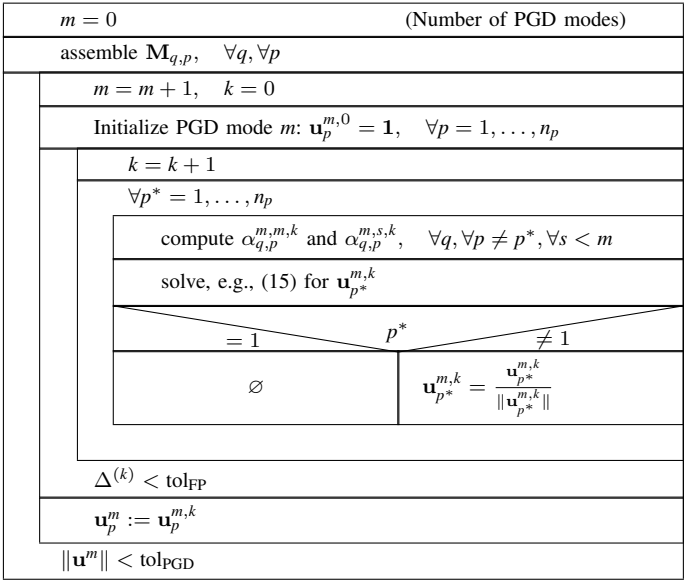}
    \caption{Nassi-Shneiderman diagram of the PGD algorithm applied to the electrokinetic subproblem.}
    \label{fig:PGDalgo}
\end{figure}

In \cref{fig:PGDalgo}, a Nassi-Shneiderman diagram of the \gls*{PGD} algorithm applied to the electrokinetic subproblem is shown.
After each parametric (non-spatial) step of the fixed point iteration, a normalization of the solution is applied. 
For the stopping criterion, the tolerances $\textrm{tol}_{\textrm{FP}}$ and $\textrm{tol}_{\textrm{PGD}}$ are used to define the accuracy of the fixed point and the \gls*{PGD} iteration, respectively.
Here, the solution change ${\Delta^{(k)}=\prod_{p=1}^{n_p}\lVert\mathbf{u}_{p}^{m,k}-\mathbf{u}_{p}^{m,k-1}\rVert}$ and the solution's magnitude $\mathbf{u}^{m}:=\prod_{p=1}^{n_{p}}\lVert\mathbf{u}_{p}^{m}\rVert$ of mode $m$ is used.

\subsection{Electrothermal Coupling}\label{sec:pgdthermal}
Once the \gls*{PGD} solution $\ensuremath{\varphi}\xspace_{\text{PGD}}^{m}$ of the electric potential is obtained, it can be used to couple to the thermal problem via the resistive losses $Q$.
The weak form of the thermostatic problem reads: find $T\in V_{\Theta}$ s.t.
\begin{align}
    \langle\lambda(\tilde{\mathbf{x}})\nabla_{\!\mathbf{x}}T(\tilde{\mathbf{x}}),\nabla_{\!\mathbf{x}}v(\tilde{\mathbf{x}})\rangle_{V_{\Theta}}
    &=\underbrace{\langle Q(\sigma,\ensuremath{\varphi}\xspace),v(\tilde{\mathbf{x}})\rangle_{V_{\Theta}}}_{\mathcal{R}},
    \label{eq:thermal_weak}
\end{align}
for all $v\in V_{\Theta}$.
The left-hand side of~\cref{eq:thermal_weak} is mathematically identical to that of~\cref{eq:estatic_weak}.
Therefore, we focus on the coupling term $\mathcal{R}$ on the right-hand side.
Again, we omit the parametric dependencies and arrive at a separated form of $\mathcal{R}$ given by
\begin{align}
    \mathcal{R}
        &=\sum_{q=1}^{n_{\sigma}}\left\langle\sigma_{q}\left|\sum_{s=1}^{m}\nabla_{\!\mathbf{x}}u^{s}\right|^{2},v\right\rangle_{V_{\Theta}}\notag\\
        &=\sum_{q=1}^{n_{\sigma}}\sum_{i=1}^{m}\sum_{j=1}^{m}\left\langle\sigma_{q}\nabla_{\!\mathbf{x}}u^{i}\cdot\nabla_{\!\mathbf{x}}u^{j},v\right\rangle_{V_{\Theta}}\notag\\
        &=\sum_{q=1}^{n_{\sigma}}\sum_{i=1}^{m}\sum_{j=1}^{m}\prod_{p=1}^{n_{p}}Q_{q,p}^{i,j},
\end{align}
with the trilinear form
\begin{equation}
    Q_{q,p}^{i,j}=\left\langle\sigma_{q,p}\nabla_{\!\mathbf{x}}u_{p}^{i},\nabla_{\!\mathbf{x}}u_{p}^{j},v_{p}\right\rangle_{V_{p}}.
\end{equation}
Note that we abused notation by using $m$ here as the number of modes of the electric problem.
To finally solve the thermal subproblem, the term $\mathcal{R}$ needs to be added to the right-hand side of the thermal equivalent to~\cref{eq:estatic_pgd} resulting in a discretized linear system similar to \cref{eq:PGDsystemDisc}.
Since we consider only a one-directional coupling, the final electrothermal solution is given by the solutions of the two subproblems.

\subsection{Using existing 3D solvers}
\label{sec:fitPGD}

Although the \gls*{PGD} is commonly based on the weak formulation, as done in \cref{sec:pgd}, discretization schemes that do no require a weak formulation can be used as well.
In each step of the fixed point iteration, either the spatial or a parametric problem needs to be solved.
Solving a parametric problem is specific to \gls*{PGD} and is typically implemented by the \gls*{PGD} tool itself.
On the other hand, the implementation of a spatial solver is more involved, motivating the usage of existing solvers.
However, a completely non-intrusive implementation of the \gls*{PGD} algorithm may not be possible, since the coefficients $\alpha_{q,p}^{m,m,k}$ and $\alpha_{q,p}^{m,s,k}$, that are used for the parametric solver, require information from the spatial solver.
Solvers that provide access to the mathematical formulation of the problem enable a straightforward computation of these coefficients.
Black box solvers may not give access to the required quantities.
Here, we use an in-house electrothermal solver based on the \gls*{FIT}~\cite{Weiland_1996aa}.
 \section{Numerical Results}
\label{sec:results}

In this Section, the \gls*{PGD} algorithm for electrothermal problems, as presented in \cref{sec:pgd}, is applied to an electrothermal 1D model problem and to the electric simulation of a 3D microelectronic chip package.
For all examples, ${\text{tol}_{\text{FP}}=\num{1e-7}}$ was used as the tolerance for the fixed point iteration.

\subsection{1D Electrothermal Model Problem}
\label{sec:1D}

\begin{figure}
    \begin{subfigure}[t]{0.49\columnwidth}
        \includegraphics{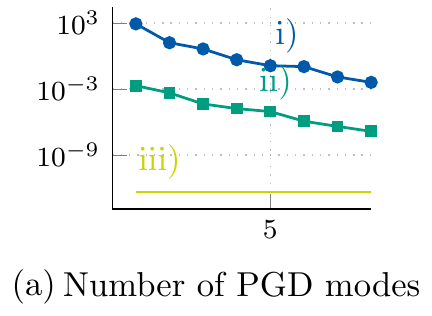}
        \phantomsubcaption\label{fig:FITterror}
    \end{subfigure}
    \hspace{-1em}
    \begin{subfigure}[t]{0.49\columnwidth}
        \includegraphics{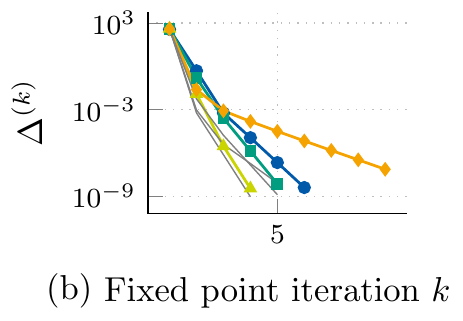}
        \phantomsubcaption\label{fig:FITtconv}
    \end{subfigure}
    \vspace{-1.5ex}
    \caption{(a) i) Magnitude $\lVert\mathbf{u}^{s}(\tilde{\mathbf{x}})\rVert$, ii) relative error of different thermal \gls*{PGD} solutions $\ensuremath{\mathbf{T}}\xspace^{m}(\tilde{\mathbf{x}})$, and iii) the spatial discretization error. (b)~Solution change $\Delta^{(k)}$ for each step of the fixed point iteration for all thermal \gls*{PGD} mode calculations.}
    \label{fig:results}
\end{figure}

We consider a 1D problem in the domain $\Omega=(0,2L)$, with $L=1$, for which \cref{eq:etspaceonly} simplifies to
\begin{subequations}
    \begin{alignat}{2}
        -\partial_{x}\left(\sigma(x;\mu) \partial_{x}{}\ensuremath{\varphi}\xspace(x;\mu)\right)&=0,&&\quad{}x\in\Omega,\\
        -\partial_{x}\left(\lambda(x;\mu)\partial_{x}{}T(x;\mu)\right)  &=\sigma(x;\mu)\left|\partial_{x}\ensuremath{\varphi}\xspace\right|^{2},&&\quad{}x\in\Omega,
    \end{alignat}
    with the boundary conditions
    \begin{alignat}{2}
        \ensuremath{\varphi}\xspace(0;\mu)&=0,\quad\ensuremath{\varphi}\xspace(2L;\mu)&&=1,\\
        T(0;\mu)&=20, \quad T(2L;\mu)&&=20,
    \end{alignat}
    \label{eq:modelproblem1D}
\end{subequations}
and with the piecewise constant material coefficients
\begin{align}
    \sigma(x;\mu)=\lambda(x;\mu)=\left\{
    \begin{aligned}
        &\mu,&&\quad\text{for }x\in(0,L),\\
        &1,  &&\quad\text{for }x\in(L,2L).\\
    \end{aligned}
    \right.
\end{align}
Problem \cref{eq:modelproblem1D} is discretized using the \gls*{FIT} with $200$ elements in $x$-direction.
Applying the \gls*{PGD} algorithm of \cref{sec:pgd} to \cref{eq:modelproblem1D}, \cref{eq:LqmFixedPoint} simplifies for the electrokinetic part with $q=1$ and $p^{*}=1$ to
\begin{equation}
    \mathcal{L}(1,m)=\int_{(0,L)}\nabla_{\!\mathbf{x}}u_{1}^{m}\cdot\nabla_{\!\mathbf{x}}v_{1}\ \mathrm{d}{}x\int_{I_{1}}\mu\left(u_{2}^{m}\right)^{2}\ \mathrm{d}\mu.
\end{equation}
Using ${I_{1}=[0.2,1]}$, $100$ points in $\mu$-direction and a tolerance of $\text{tol}_{\text{PGD}}=\num{1e-2}$, an electrothermal \gls*{PGD} solution of $3$ electric and $8$ thermal modes is obtained.
The relative error of the thermal solution of the resulting electrothermal surrogate model with respect to the analytical solution is shown in \cref{fig:FITterror}, curve~ii).
Additionally, $\lVert{}u^{s}\rVert$ and the discretization error of the \gls*{FIT} model are shown.
In \cref{fig:FITtconv}, the convergence of the fixed point iteration for the individual modes of the thermal solution is observed by plotting the solution change $\Delta^{(k)}$.
In total, only $11$ electrothermal \gls*{PGD} modes and $42$ spatial field solver calls are required to obtain an accurate model for a wide parameter range.

\subsection{Chip problem}\label{sec:chip}
\begin{figure}
    \centering
        \includegraphics{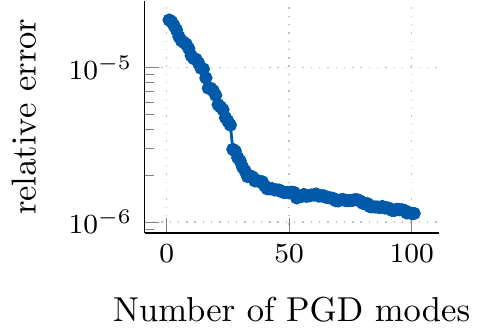}
    \caption{Relative error of the electric solution of the chip model with respect to the standard FIT solution.}
    \label{fig:chipError}
\end{figure}

\begin{figure}
    \centering
    \begin{subfigure}{0.24\columnwidth}
        \centering
        \includegraphics{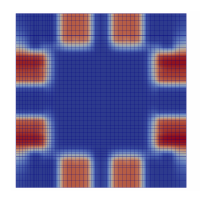}
                \caption{$\mathbf{u}_{1}^{1}(\mathbf{x})$}
    \end{subfigure}
    \begin{subfigure}{0.24\columnwidth}
        \centering
        \includegraphics{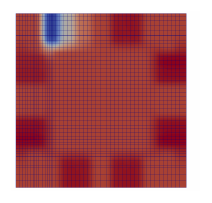}
                \caption{$\mathbf{u}_{1}^{2}(\mathbf{x})$}
    \end{subfigure}
    \begin{subfigure}{0.24\columnwidth}
        \centering
        \includegraphics{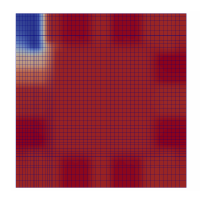}
                \caption{$\mathbf{u}_{1}^{3}(\mathbf{x})$}
    \end{subfigure}
    \begin{subfigure}{0.24\columnwidth}
        \centering
        \includegraphics{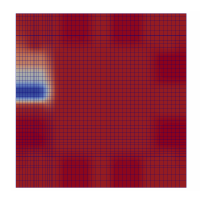}
                \caption{$\mathbf{u}_{1}^{4}(\mathbf{x})$}
    \end{subfigure}
    \caption{First four spatial modes of the chip's electric solution.}
    \label{fig:chipResultsSpatial}
\end{figure}

A microelectronic chip package geometry~\cite{Casper_2016aa} with 12 internal bond wires of uncertain length is considered, see \cref{fig:chip}.
The wires are modeled by an uncertain length $\overline{L}\in[10^{-4},10^{-3}]$ discretized by \num{100} points.
Following~\cite{Casper_2018ah}, we model the wires using a 1D-3D coupling approach with zero coupling radius and two 1D points each.
In \cref{fig:chipError}, the relative error of the electric \gls*{PGD} solution with respect to the standard FIT solution is shown.
The qualitative distribution for the first four spatial modes of the solution is shown in \cref{fig:chipResultsSpatial}.
Whereas the first mode is the main representative of the global field distribution, the following modes shows the local influence of the uncertain wires.
To obtain an accuracy of $\num{1.14e-6}$, $100$ modes using $326$ spatial field solver calls were required.

 \section{Conclusion}
\label{sec:conclusion}

A \gls*{PGD} formulation for static electrothermal problems including an arbitrary number of uncertain parameters has been presented.
In contrast to other techniques, the \gls*{PGD} avoids exponential dependence of the computational cost on the number of uncertain parameters and thus, it circumvents the curse of dimensionality.
The electrothermal coupling term requires a non-standard trilinear form modeling the resistive losses.
For numerical validation, a 1D model problem and a 3D microelectronic chip package example were considered.
Extending the algorithm to a transient bi-directional electrothermal coupling and the computation of multi-parameter examples are subject to future work.
Using the \gls*{PGD} approach for optimization problems could be another possible application.
 
\section*{Acknowledgment}
This work is supported by the ``Excellence Initiative" of the German Federal and State Governments and by the Graduate School of Computational Engineering at Technische Universität Darmstadt.

\clearpage
 
\end{document}